\documentstyle[preprint,revtex]{aps}
\begin{document}
\draft

\def\lsim{\mathrel{\lower2.5pt\vbox{\lineskip=0pt\baselineskip=0pt
           \hbox{$<$}\hbox{$\sim$}}}}
\def\gsim{\mathrel{\lower2.5pt\vbox{\lineskip=0pt\baselineskip=0pt
           \hbox{$>$}\hbox{$\sim$}}}}

\rightline{{\bf IEM--FT--55/92}}
\rightline{{\bf April 1992}}
\vskip16cm
Submitted to {\em Physical Review Letters}.

\vskip-14cm

\begin{title}
{\bf Axions with variable masses}
\end{title}

\author{Juan Garc\'{\i}a--Bellido}

\begin{instit}
Instituto de Estructura de la Materia,\
CSIC,\ \ Serrano\ 123,\ \ E--28006\ \ Madrid,\ \ Spain
\end{instit}

\begin{abstract}
Axions with variable masses, in the context of a scalar--tensor
gravity theory, give a large entropy production during the
matter era. The subsequent axion dilution is proportional to
their present energy density. Depending on the parameters
($\beta_I,\beta_V$) of the model, this dilution relaxes or even
eludes the cosmological bound on the axion mass, therefore
opening the so--called ``axion window".
\end{abstract}


\narrowtext

The Peccei--Quinn mechanism \cite{PQM} is the most elegant
solution to the strong CP--problem \cite{SCP}. The axion
\cite{AXI} is the pseudo--Goldstone boson associated with the
spontaneous symmetry breaking of the global $U(1)_{\rm PQ}$.
Weakly interacting axions, coming from a large symmetry breaking
scale, are called ``invisible" \cite{INV} and arise
naturally in superstring models.

There is no theoretical prediction on the mass of the axion.
There are, however, strong phenomenological constraints on $m_a$
coming from astrophysics and cosmology \cite{REV}. The
astrophysical bounds arise from energy--loss rates in SN1987A,
$m_a \lsim 10^{-2}$--$10^{-3}$ eV \cite{AST}, while the
cosmological constraints come from the axion contribution to
the critical density of the universe. Axions produced by the
``misalignment" mechanism \cite{REV}, associated with coherent
oscillations of the axion field, contribute with \cite{COS}
\begin{equation}
\label{OMA}
\Omega_a^{\rm OSC} h^2 \simeq\ \theta_1^2
\left(\frac{m_a}{10^{-5}\ {\rm eV}}\right)^{-1.18},
\end{equation}
where $h$ is Hubble's constant in units of 100 km s$^{-1}$
Mpc$^{-1}$ and $\theta_1$ is the initial ``misalignment"
angle at horizon crossing, randomly distributed between $-\pi$
and $\pi$ \cite{MST}. Moreover, cosmic strings are produced
in the spontaneous breaking of the global $U(1)_{\rm PQ}$, which
decay by radiating axions \cite{STR,TRS}. The contribution of
these axions to the critical density can be estimated as
\cite{STR} (see however ref.\cite{TRS})
\begin{equation}
\label{OST}
\Omega_a^{\rm STR} h^2 \simeq\
\left(\frac{m_a}{10^{-3}\ {\rm eV}}\right)^{-1.18}.
\end{equation}
The age of the universe imposes a cosmological constraint
on the axion energy density, $\Omega_a h^2 \lsim 1$, which
requires, see eqs.(\ref{OMA}, \ref{OST}),
$m_a \gsim 10^{-3}$--$10^{-5}$ eV.
The astrophysical upper bounds and the cosmological lower bounds
on the axion mass already close the so--called ``axion window".
Furthermore, the PQ--symmetry breaking leaves a
residual discrete $Z_N$ symmetry, leading to potentially
dangerous axionic domain walls \cite{DOM}.

In the estimation of the axion contribution to the critical
density (\ref{OMA}, \ref{OST}) it was assumed that there has
been no significant entropy production at
later stages of the evolution of the universe. If, on the other
hand, the entropy per comoving volume $S$ is increased by a factor
$\gamma$ since the time of axion production, then $\Omega_a h^2$
is reduced by the same factor \cite{KT}.
There has been several attempts to open the axion window and
simultaneously solve the axionic domain wall problem,
the most important one being the use of inflation \cite{INF}.
Other physical processes, apart from inflation, which
may be responsible for axion dilution are large entropy
production at late stages of the universe, like decaying
particles out of equilibrium and first order phase transitions
\cite{LEP}. It is important to know the different sources of
axion dilution since there is a proposal of an experimental
search for dark matter axions \cite{SIK} that will not detect an
axion with mass $m_a \ll 10^{-5}$eV, which on the other hand
could be allowed by those processes.

In this letter we analyze the entropy production and subsequent
dilution of dark matter axions with {\em variable} masses in the
context of a scalar--tensor gravity theory \cite{SCT},
with {\em different} couplings of the scalar field
to visible and dark matter \cite{DGG}. This kind of models arise
naturally from superstring theory \cite{NPB,CQG} and have been
considered in the context of extended inflation \cite{EI}. They
explicitly violate the weak equivalence principle but are not
ruled out by observations \cite{DGG,CQG}. The action of such a
model can be written in the Einstein frame as
\begin{equation}
\label{SMD}
\begin{array}{rl}
S=&{\displaystyle
\int d^4x \sqrt{-g} \left( R - {\textstyle \frac{1}{2}}
(\partial\phi)^2
+ 16\pi e^{\beta_I\phi} {\cal L}_{m_I} \right.} \vspace{1mm} \\
&{\displaystyle \left. \hspace{3cm}
+ 16\pi e^{\beta_V\phi} {\cal L}_{m_V} \right) },
\end{array}
\end{equation}
where $(\beta_I,\beta_V)$ parametrize the scalar couplings to
invisible ({\em i.e.} axionic) and visible ({\em i.e.} baryonic)
matter sectors. There is no theoretical prediction on the value
of these parameters but there are bounds \cite{DGG,CQG,TDG}
coming from radar time--delay experiments \cite{RAD,SCT},
the age of the universe \cite{AGE,KT} and primordial
nucleosynthesis \cite{KAO,PLB}
\begin{equation}
\label{COT}
\begin{array}{l}
\beta_V < 0.022 \\
\beta_I < 0.674 \\
\beta_I \beta_V < 3\times 10^{-4}.
\end{array}
\end{equation}

During the radiation era all scalar--tensor theories behave like
general relativity, since the scalar field is then constant
\cite{NPB}. Therefore, the usual mechanisms of axion production
are not modified. At low temperatures, the axion
behaves like ordinary non--relativistic matter with
energy density
\begin{equation}
\label{RHO}
\rho_a = n_a\ m_a = \frac{N_a\ m_a}{R^3},
\end{equation}
where $N_a=n_a R^3$ is the conserved number of dark matter
axions \cite{CQG}. Suppose that axions with {\em variable} masses
dominate the evolution of the universe during the matter era.
The scalar coupling to the axion mass will be responsible for a
large entropy production, which can dilute the axion
contribution to the critical density. Let us
consider a perfect fluid composed of dark matter axions with
variable masse. The energy--momentum tensor, in the conformal
frame of constant masses for visible matter, satisfies the
conservation equation \cite{NPB,CQG}
\begin{equation}
\label{EMC}
T^{\mu\nu}_{\ \ ;\nu} = (\beta_I-\beta_V) \partial^{\mu}\phi
\ T^\lambda_{\ \lambda} .
\end{equation}
The total amount of entropy production due to the coupling of
the scalar field can be computed by writing the energy
conservation equation (\ref{EMC}) in a Robertson--Walker frame
\begin{equation}
\label{EDC}
\frac{d}{dt}(\rho R^3) + p \frac{d}{dt}(R^3) =
\frac{1}{m}\frac{d m}{dt} (\rho-3p) R^3
\end{equation}
and comparing it with the second law of Thermodynamics
$dU + pdV = TdS$. For non--relativistic axions (\ref{RHO})
we obtain
\begin{equation}
\label{TDS}
TdS \simeq\ N_a\ d m_a(\phi).
\end{equation}
The total entropy increase per comoving volume from the
time of equal matter and radiation energy density to now
is given by \cite{CQG}
\begin{equation}
\label{DS}
\begin{array}{rl}
{\displaystyle
\Delta S = \int_{t_{eq}}^{t_o} \frac{N dm}{T} }
&{\displaystyle
\simeq \frac{2\beta_I(\beta_V - \beta_I)}{1-4\beta_I^2}
\ \frac{N_a m_a(t_o)}{T_o} }\vspace{2mm}\\
&{\displaystyle
\equiv\ k(\beta_I,\beta_V) \ \frac{N_a m_a}{T_o} } ,
\end{array}
\end{equation}
where $T_o\sim 1.4$ K is the axion temperature today \cite{COS}.
Using the fact that baryons are non--relativistic,
$\rho_B = n_B\ m_B$, and their contribution to the critical
density is $\Omega_B h^2 \sim 10^{-2}$ \cite{KT}, we find
\begin{equation}
\label{RDS}
\gamma - 1 = \frac{\Delta S}{S} \simeq \ 10^2\ k(\beta)
\ \Omega_a h^2\ \frac{\eta N_\gamma}{S}\ \frac{m_B}{T_o},
\end{equation}
where $\eta=n_B/n_\gamma \sim 4\times 10^{-10}$ is the baryon to
photon ratio, $m_B\sim 1$ GeV is the proton mass and the total
entropy per comoving volume of the universe is related to the
number of photons by $S\simeq 7 N_\gamma$ \cite{KT}. Note that
the entropy production is proportional to the axion energy
density (\ref{OMA}, \ref{OST}). Therefore, the fraction of
critical density contributed by axions with variable masses
($\Omega_a^{\rm VAR} = \Omega_a /\gamma$) can be bounded as
\begin{equation}
\label{CRD}
\Omega_a^{\rm VAR} h^2 \simeq \frac{\Omega_a h^2}
{1 + 5\times 10^4\ k(\beta)\ \Omega_a h^2} \lsim 1.
\end{equation}
{}From (\ref{COT}) we cannot deduce a bound on $k(\beta)>0$.
For $k(\beta_I,\beta_V)\lsim 2\times 10^{-5}$ the mass of the
axion satifies
\begin{equation}
\label{BOM}
m_a \gsim 10^{-5}\ {\rm eV}\ \left(1 - 5\times 10^4
\ k(\beta) \right)^{0.85},
\end{equation}
which relaxes the cosmological bound on $m_a$ \cite{COS}. On
the other hand, for $k(\beta_I,\beta_V)\gsim 2\times 10^{-5}$,
we find {\em no constraint} on the axion mass but only on the
parameters of our model (\ref{SMD}). Using the previous bounds
(\ref{COT}) we estimate
\begin{equation}
\label{CON}
\beta_I < 0.017,
\end{equation}
which improves significantly the bound on $\beta_I$, see
eq.(\ref{COT}).

In conclusion, we believe that axions with variable masses are
very plausible candidates for cold dark matter. They could be
responsible for the halo of spiral galaxies and at the same time
provide closure density. Entropy production due to the scalar
coupling to the axion mass is a simple alternative mechanism to
inflation for opening the ``axion window". Note that, contrary
to inflation, this is a selective mechanism which does not
produce any baryon dilution.

It is a pleasure to thank Mariano Quir\'os and Alberto Casas for a
careful reading of the manuscript and useful suggestions. This
work was supported by CICYT under contract No. AEN90--0139 and
by a MEC--FPI Grant.

\end{document}